\newcommand{\be}{\begin{eqnarray}}
\newcommand{\ee}{\end{eqnarray}}

\documentclass[twocolumn,aps,showpacs]{revtex4}
\usepackage{graphics}
\begin{document}
\title{Proton and pion transverse spectra at RHIC from radial flow and
  finite size effects} 
\author{Alejandro Ayala$^\dagger$, Eleazar Cuautle$^\dagger$, J.
Magnin$^\ddagger$, Luis Manuel Monta\~no$^*$}   
\affiliation{$^\dagger$Instituto de Ciencias Nucleares, Universidad
Nacional Aut\'onoma de M\'exico, Apartado Postal 70-543, M\'exico
Distrito Federal 04510, Mexico.\\
$^\ddagger$Centro Brasileiro de Pesquisas Fisicas, Rua Dr. Xavier
Sigaud 150-Urca CEP 22290-180, Rio de Janeiro, Brazil.\\
$^*$Centro de Investigaci\'on y de Estudios Avanzados del IPN,
Apartado Postal 14-740,
M\'exico Distrito Federal 07000, Mexico}

\begin{abstract}

We show that the proton and pion transverse momentum distributions
measured at RHIC for all collision centralities for pions and most of the
collision centralities for protons, can be simultaneously
described in terms of a thermal model with common values for the radial 
flow and temperature, when accounting for the finite size of the interaction
region at the time of decoupling. We show that this description is obtained in
terms of a simple scaling law of the size of the interaction region with the
number of participants in the collision. The behavior of the proton to pion
ratio at mid-rapidity can also be understood as a consequence of the strength
of the radial flow and system size reached at RHIC energies.

\end{abstract}

\pacs{25.75.-q}

\maketitle

\date{\today}

\section{Introduction}

The unexpected behavior of the proton to pion ratio as a function of $p_t$ has
been taken as an indication of the onset of the thermal recombination of
quarks as an important mechanism for hadron production at RHIC
energies in Au + Au collisions~\cite{PHENIXBM}. In its simplest form, thermal
recombination invokes a densely populated parton phase space to allow the
statistical formation of hadrons from constituent quarks assigning degeneracy
factors appropriate for either mesons or baryons~\cite{recomb}. 

An important short come of the recombination scenario is that it ignores
inelastic and elastic scattering experienced by hadrons before kinetic 
freeze-out and thus that neither particle abundance nor their
spectra are fixed right after recombination. A more appropriate description of
statistical systems, that includes the fact that the detailed history is
washed out by means of interactions after a long enough time, can be given in
terms of global features which survive all the way to the end of the system's
evolution. One of these global features is flow, in particular radial
flow.

It has been observed that the magnitude of the radial flow
velocity exhibits a 50\% increase from AGS and SPS to RHIC
energies~\cite{Tserruya}. Recall that in $p+p$ collisions, where no effects of
radial flow exist, it is known that the proton to pion ratio as a
function of $p_t$ remains basically unchanged, never exceeding one, for
collision energies ranging from 19.4 GeV at the Tevatron, 44.6 and 52.8 GeV at
ISR up to 200 GeV at RHIC~\cite{STAR}. In contrast, the proton to pion ratio
in Au + Au collisions at RHIC reaches and even exceeds one for $p_t\sim 2$
GeV. Therefore, if recombination of thermal partons has anything to do with
this behavior it is clear that a thermal description of the individual
particle spectra has to be possible at least up to such $p_t$
values. Nevertheless, in Ref.~\cite{PHENIXBM}, a fit to a thermal
model that attempts a description of particle spectra up to $p_t\sim
2$ GeV in terms of an intrinsic freeze-out temperature $T_0$, together
with radial flow, yields values of order $T_0\sim 180$ MeV,
which is closer to the hadronization temperature than to the kinetic
freeze-out temperature.

Yet another intriguing behavior that concerns freeze-out temperatures and
expansion velocities in thermal models is their relation as a function of the
centrality of the collisions, which, within the usual thermal model
calculations, can be stated as an increase in flow
together with a decrease in temperature as the centrality of the collision
increases. This behavior is usually attributed to the larger time spent by
the system in the hadronic phase for the most central collisions allowing for
the development of flow and consequently decreasing the values for the kinetic
freeze-out temperature~\cite{Hung,ppr}. However, as results from elliptic flow
analyses seem to indicate~\cite{flow}, flow is generated early, in the
partonic phase of the collision. Moreover,
kinetic freeze-out temperatures can also be thought of as a global feature of
strongly interacting systems that reflect the average kinetic energy needed
for the system to decouple. In fact, a systematic study of HBT data and
particle yields for pions at mid-rapidity from AGS to RHIC
energies~\cite{CERES} shows that this average energy is independent of
centrality and beam energy. Therefore one can ask if an alternative 
description, with common values of temperature and flow velocity, reflecting
the above property of strongly interacting systems, can be achieved for all
centralities. As we show, the key ingredient that allows a description of
particle spectra in a thermal model, including radial flow, and that addresses
the above mentioned phenomena from a unifying point of view, is the
realization that particle production and successive freeze-out in a
relativistic heavy-ion environment takes place during small time scales, of
order 10 fm, and consequently within small volumes.  

Although not commonly considered, small size
effects are important in the description of a variety of phenomena associated
with statistical systems such as the late-stage growth of nucleated bubbles
during a first order phase transition~\cite{Raju} and the statistical
hadronization model~\cite{Becattini}. Finite size effects are also known to
influence the interpretation of the correlation lengths in Hanbury Brown-Twiss
analysis in the context of relativistic heavy-ion collisions~\cite{Zhang,
Ayalaint}. 

Recall that useful microscopical information in this kind of
collisions can be obtained from comparing the average interparticle
separation during the collision evolution to the range of strong
interactions. For the case of pions (the most copiously produced
particles in the collision) right after the collision, the system   
is better described as a liquid rather than as a gas~\cite{Shuryak}.
One important consequence is the appearance of
a surface tension which acts as a reflecting boundary
for the particles that move toward it. The reflection
details depend on the wave length of the incident particle
but the important property introduced by the reflecting
surface is that it allows very little wave function
leakage and, to a good approximation, the wave
functions vanish outside this boundary. When the
average separation of the particles in the system becomes larger
than the range of the strong interaction, they become
a free gas but because of the short interaction range,
the transition between the liquid and the gas stages
is very rapid and the momentum distribution is determined
by the distribution just before freeze-out.

To be concrete, we need to compare the pion separation $d$
to the average range of the pion strong interaction ($d_s\sim 1.4$ fm). For
typically accepted values for the density and formation times~\cite{ABM}, it is
possible to show that $d\sim 0.6$ fm $<\ d_s$ and the condition to regard the
pion system as a liquid is met.  

Qualitatively, the behavior of thermal particle spectra including finite
size effects deviates from a simple exponential fall-off at high 
momentum since from the Heisenberg uncertainty principle, the more
localized the states are in coordinate space, the wider their
spread will be in momentum space. In terms of the discrete set of energy
states describing the particle system, this behavior can be understood as
arising from a higher density of states at large energy as compared to
a calculation without finite size effects. These ideas have been applied to
the description of charged and neutral pion spectra measured at RHIC with a
good agreement for the transverse momentum interval
$0<p_t\lesssim3$~\cite{Ayalanew}. 

In this paper we compute the transverse momentum distribution for pions and
extend the above ideas to also include protons in the description,
assuming thermal equilibrium together with radial flow and 
accounting for finite size effects at decoupling. By comparing to data
on pion and proton spectra on Au+Au collisions at $\sqrt{s_{NN}}=200$
GeV~\cite{PHENIXBM}, we show that for temperatures and collective transverse
flow within values corresponding to kinetic freeze-out conditions, the
transverse momentum distributions can be described with common values of
temperature and expansion velocities for all collision centralities for pions
and for most of the collision centralities for protons.

The work is organized as follows: In Sec.~\ref{II} we present the
basics of the model to compute pion and proton distributions. In
Sec.~\ref{III} we compute the transverse momentum distributions for
protons and pions comparing to data on Au+Au collisions at $\sqrt{s_{NN}}=200$
GeV. We show that a good agreement with these data for
different collision centralities can be achieved by assuming a simple
scaling of the radii with the cube root of the number of participants in the
collision. In Sec.~\ref{IV} we compute the pion correlation
function and also extract the size of the system as a function of the
cube root of the number of participants in the collision. We finally
conclude in Sec.~\ref{concl}.

\section{The model}\label{II}

We consider a scenario where finite size effects are included by restricting
the system of particles to be confined within a
volume of the size of the fireball at freeze-out. Since we aim to describe
spectra at central rapidities, it suffices to 
take the confining volume as a sphere of radius $R$ (fireball) as viewed from
the center of mass of the colliding nuclei at the time of
decoupling~\cite{Ayalanoex}. This time needs not be the same over the entire
reaction volume. Nevertheless, in the spirit of the fireball model we consider
that decoupling takes place over a constant time surface in space-time. This
assumption should be essentially correct if the freeze-out interval is short
compared to the system's life time. Though some of the particles emitted in
the central rapidity region could originate from a finite range of
longitudinal positions due to thermal smearing, we will consider that most of
the central rapidity particles come from the central 
spatial region and thus neglect possible effects on these particles from a
different longitudinal and transverse expansion velocities.

In the case of bosons, the wave functions that incorporate the effects of a
finite size system have been found in Ref.~\cite{Ayalanoex}, where we refer
the reader for further details of the model. These wave functions are given as
the stationary solutions of the wave equation for bosons, namely the
Klein-Gordon equation 
\be
   \left(\frac{\partial^2}{\partial t^2} -
   \nabla^2 + m^2\right)
   \phi({\mathbf r},t)=0
   \label{eq:neweq}
\ee
subject to the boundary condition
$
   \phi(|{\mathbf r}|=R,t)=0\, ,
$
and finite at the origin. The normalized stationary states are
\be
   \phi_{nlm'}({\mathbf r},t)&=&\frac{e^{-iE_{nl}t}}{R\ J_{l+3/2}(k_{nl}R)}
   \nonumber\\
   &\times&\frac{Y_{lm'}({\mathbf{\hat{r}}})J_{l+1/2}(k_{nl}r)}
   {\sqrt{rE_{nl}}}\, .
   \label{eq:solnew}
\ee
In the case of fermions, the wave functions are found as the stationary
solutions of the Dirac equation
\be
   \left(i\gamma^\mu\partial_\mu - m\right)\psi({\mathbf r},t)=0
   \label{dirac}
\ee
subject to the the boundary condition 
$
   \psi(|{\mathbf r}|=R,t)=0\, ,
$
and also finite at the origin. It is easy to show that the normalized
stationary states are  
\be
   \psi_{nlm'}({\mathbf r},t)&=&
   \frac{e^{-iE_{nl}t}}
   {2mR\ J_{l+3/2}(k_{nl}R)}
   \nonumber\\
   &\times&
   \left\{\begin{array}{r}
               E_{nl} + m + i{\mbox{\boldmath${\sigma}$}}\cdot
               {\mbox{\boldmath${\nabla}$}}\\
               -E_{nl} + m - i{\mbox{\boldmath${\sigma}$}}\cdot
               {\mbox{\boldmath${\nabla}$}}
           \end{array}\right\}\nonumber\\
   &\times&\frac{Y_{lm'}({\mathbf{\hat{r}}})J_{l+1/2}(k_{nl}r)}{\sqrt{r}}\, .
   \label{eq:soldir}
\ee
In Eqs.~(\ref{eq:solnew}) and~(\ref{eq:soldir}), 
$J_\nu$ is a Bessel function of the first kind and $Y_{lm'}$ is a 
spherical harmonic, {\boldmath$\sigma$} are the Pauli matrices and
the parameters $k_{nl}$ are related to the energy  
eigenvalues $E_{nl}$ by
$
   E_{nl}^2=k_{nl}^2+m^2
$
and are given as the solutions to
$
   J_{l+1/2}(k_{nl}R)=0\, .
$
The contribution to the thermal invariant distribution from a
state with quantum numbers $\{n,l,m'\}$ is given by
\be
   E\frac{d^3N_{nlm'}}{d^3p}=
   \int \frac{d\Sigma}{(2\pi)^3}(k_{nl}\cdot
   u) f(k_{nl}\cdot v){\mathcal{W}}_{nlm'}({\mathbf{p}},{\mathbf{r}}),
   \label{distone}
\ee
where ${\mathcal{W}}_{nlm'}({\mathbf{p}},{\mathbf{r}})$ is the Wigner
transform and $f(k_{nl}\cdot v)$ the thermal occupation factor of the
state, respectively. The four-vectors $v^\mu$ and $u^\mu$ 
represent the collective flow four-velocity and a four-vector of
magnitude one, normal to the freeze-out hypersurface $\Sigma$,
respectively.

In order to consider a situation where freeze-out happens at a
fixed time and within a spherical volume of radius $R$, the unit
four-vector $u^\mu$ can be chosen as $u^\mu=(1,{\mathbf 0})$. 
To keep matters simple, we also consider a thermal occupation factor of
the Maxwell-Boltzmann kind
$
   f(k_{nl}\cdot v)=e^{-k_{nl}\cdot v/T}
$
where $T$ is the system's temperature. The four-vector $v^\mu$ is
parametrized as
$
   v^\mu=\gamma(1,{\mathbf v}),
$
and we choose a radial profile for the vector ${\mathbf v}$ such as
$
   {\mathbf v}=\beta{\mathbf r}/R,
$
where the parameter $\beta$ represents the surface expansion
velocity. Correspondingly, the gamma factor is given by
\be
   \gamma=\frac{1}{\sqrt{1-\left(\beta\frac{r}{R}\right)^2}}.
   \label{gammafac}
\ee
Nonetheless, in order to continue to keep matters as simple as
possible, and be able to analytically perform the integrations in
Eq.~(\ref{distone}), we will instead consider that the gamma factor is
a constant evaluated at the average transverse expansion velocity, namely,
\be
   \gamma\rightarrow\bar{\gamma}=
   \frac{1}{\sqrt{1-(3\beta /4)^2}},
   \label{gammaaver}
\ee
where the average is computed by assuming that the matter distribution is
uniform within the fireball.

We take the four-vector $k_{nl}^\mu=(E_{nl},{\mathbf k}_{nl})$,
and choose ${\mathbf k}_{nl}\parallel{\mathbf p}$. This choice is
motivated from the continuum, boundless limit, where the
relativistically invariant exponent in the thermal occupation factor
becomes $\gamma(E-{\mathbf p}\cdot{\mathbf v})$.

\begin{figure}[t!] 
{\centering
\resizebox*{0.4\textwidth}
{0.34\textheight}{\includegraphics{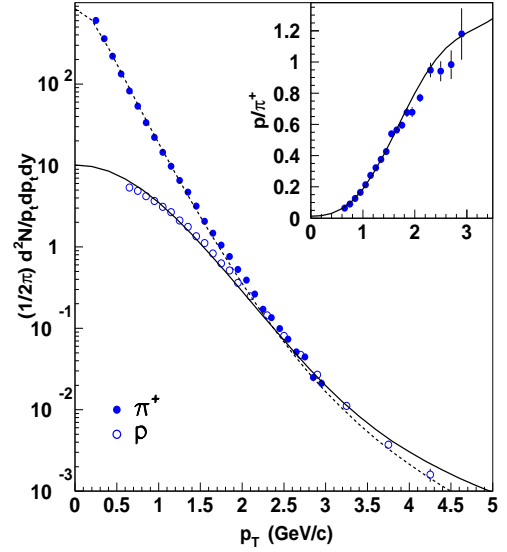}}
\par}
\caption{(Color online) Invariant $\pi^+$ and $p$ distributions as a
function of $p_t$ for 
$T=117$ MeV, $R=8$ fm and $\beta_\pi=0.6$ and $\beta_p=0.53$. Also shown in
the insert is the ratio $p/\pi^+$ of these distributions. Data are from
Ref.~\cite{PHENIXBM} for collisions with $0-10\%$ of centrality.}  
\label{fig1}
\end{figure}

Summing over all the states, the invariant thermal distribution for bosons and
fermions are given by 
\be
   E\frac{d^3N_{\mbox{\small{b}}}}{d^3p}&=&
   {\mathcal {N}}_{\mbox{\small{b}}}\sum_{nl}
   \frac{(2l+1)}{(2\pi)}\frac{k_{nl}^2E_{nl}\ e^{-\bar{\gamma}E_{nl}/T}}
   {\sqrt{p^2+\left(\frac{\bar{\gamma}\beta
   k_{nl}}{2RT}\right)^2}}\nonumber\\
   &\times&\frac{\left|J_{l+1/2}\left(pR-i
   \frac{\bar{\gamma}\beta k_{nl}}{2T}\right)\right|^2}
   {\left[p^2-k_{nl}^2-
   \left(\frac{\bar{\gamma}\beta k_{nl}}{2RT}\right)^2\right]^2 
   + \left[\frac{\bar{\gamma}\beta p k_{nl}}{RT}\right]^2},
   \label{invdistallbos}
\ee
\be
   E\frac{d^3N_{\mbox{\small{f}}}}{d^3p}&=&
   {\mathcal {N}}_{\mbox{\small{f}}}\sum_{nl}
   \frac{(2l+1)}{(2\pi)}\nonumber\\
   &\times&
   \left[\frac{(E_{nl}^2 + m^2 + p^2) + (\beta k_{nl}/2RT)^2}{m^2}
   \right]\nonumber\\
   &\times&
   \frac{k_{nl}^2E_{nl}\ e^{-\bar{\gamma}E_{nl}/T}}
   {\sqrt{p^2+\left(\frac{\bar{\gamma}\beta
   k_{nl}}{2RT}\right)^2}}\nonumber\\
   &\times&\frac{\left|J_{l+1/2}\left(pR-i
   \frac{\bar{\gamma}\beta k_{nl}}{2T}\right)\right|^2}
   {\left[p^2-k_{nl}^2-
   \left(\frac{\bar{\gamma}\beta k_{nl}}{2RT}\right)^2\right]^2 
   + \left[\frac{\bar{\gamma}\beta p k_{nl}}{RT}\right]^2},
   \label{invdistallfer}
\ee
respectively. The factor $(2l+1)$ in Eqs.~(\ref{invdistallbos})
and~(\ref{invdistallfer}) comes from the degeneracy of a state with a given
angular momentum eigenvalue $l$. ${\mathcal N}_{\mbox{\small{b}}}$ and
${\mathcal N}_{\mbox{\small{f}}}$ are normalization constants.


\begin{figure}[t!] 
{\centering
\resizebox*{0.5\textwidth}
{0.34\textheight}{\includegraphics{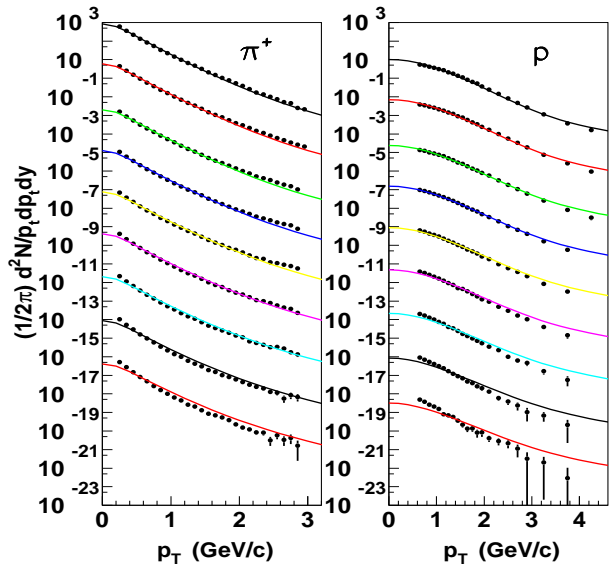}}
\par}
\caption{(Color online) Invariant $\pi^+$ (left panel) and $p$ (right
  panel) distributions 
  as a function of $p_t$ for different collision centralities. The uppermost
  curves correspond to the most central collisions, $0-10\%$ with the
  centrality decreasing in intervals of $10\%$~\cite{PHENIXBM} multiplied by
  successive factors of $10^{-2}$. The size of
  the equivalent spherical region is calculated according to the relation
  $R=R_0+C(N_{\mbox{\tiny{part}}}/2)^{1/3}$, with $R_0=1$ fm and $C=1.28$. The curves
  represent the theoretical calculation with $T_{\mbox{\small{$p,\pi$}}}=117$
  MeV, $\beta_{\mbox{\small{$\pi$}}}=0.6$ and
  $\beta_{\mbox{\small{$p$}}}=0.53$, which describe the most central
  collisions data.}
\label{fig2}
\end{figure}

The contrast between a calculation with and without finite size effects, can
be appreciated by looking at Fig.~5 in Ref.~\cite{Ayalanew} were it is shown a
comparison between the invariant pion 
distribution as a function of $p_t$ computed for $T=120$ MeV, $\beta=0.6$ with
finite size effects ($R=8$ fm) and without them. The curves are also compared
to data on positive pions from PHENIX~\cite{PHENIXBM}. We
notice from the figure that the curve with finite size effects does a very
good job describing the data for all values of $p_t$ in this range. In
contrast, a calculation where no effects of a finite size are included, and
thus the wave function of a given state is simply a plane wave, does not
describe the data over the considered range when use is made of the same
values for $T$ and $\beta$ as for the case of the calculation with a confining
volume. 

\section{Transverse spectra}\label{III}

We now compare the model to data on mid-rapidity positive pions together with
protons from central Au + Au collisions at $\sqrt{s_{NN}}=200$ GeV measured in
RHIC~\cite{PHENIXBM}. We perform a $\chi^2$ fit to each
spectra. The fit parameters are the pion and proton fireball radii,
$R_{\mbox{\small{$\pi$}}}$, $R_{\mbox{\small{$p$}}}$, temperatures
$T_{\mbox{\small{$\pi$}}}$, $T_{\mbox{\small{$p$}}}$, surface radial flow
velocities $\beta_{\mbox{\small{$\pi$}}}$, $\beta_{\mbox{\small{$p$}}}$ and
normalizations ${\mathcal{N}}_{\mbox{\small{$\pi$}}}$,
${\mathcal{N}}_{\mbox{\small{$p$}}}$. Based on the success of the description
of the central rapidity pion data obtained in Ref.~\cite{Ayalanew} up to
$p_t\sim 3$ GeV, we first fix the parameters describing the pion data
with the minimization procedure. The parameters thus obtained are
$R_{\mbox{\small{$\pi$}}}=8$ fm, $T_{\mbox{\small{$\pi$}}}=117$ MeV and
$\beta_{\mbox{\small{$\pi$}}}=0.6$ which are basically the same as the ones
obtained in Ref.~\cite{Ayalanew} were only the normalization was left as a
free parameter and the rest were set to reasonable values that describe
freeze-out conditions at RHIC. 

Next, in order to find the parameters that describe the proton
spectrum, we fix the values of any two of the parameters
$R_{\mbox{\small{$p$}}}$, $T_{\mbox{\small{$p$}}}$ and
$\beta_{\mbox{\small{$p$}}}$ to be the same as the corresponding parameters
describing the pions, leaving the third parameter, along with the
normalization constant ${\mathcal{N}}_{\mbox{\small{$p$}}}$ free. The optimum
set of parameters obtained with this procedure correspond to
$R_{\mbox{\small{$p$}}}=8$ fm, $T_{\mbox{\small{$p$}}}=117$ MeV and
$\beta_{\mbox{\small{$p$}}}=0.53$. Figure~\ref{fig1} shows the distributions
for pions and protons for central collisions $(0-10\%)$~\cite{PHENIXBM}
compared to the theoretical calculation with the best parameters obtained. We
notice that the proton data are well described by the model up to $p_t\sim 3$
GeV for a temperature and system's size equal to the corresponding parameters
for the pions but that the magnitude of $\beta_{\mbox{\small{$p$}}}$ is about
10\% smaller than $\beta_{\mbox{\small{$\pi$}}}$. We recall that in order to
find an analytical expression for the momentum distributions, we resorted to
approximate the gamma factor in Eq.~(\ref{gammafac}) by the average gamma
factor in Eq.~(\ref{gammaaver}). Since the effect of the same radial
flow is stronger for particles with larger mass, it is therefore
natural to expect that with this approximation we introduce a
discrepancy in the description of the flow for particles with
different masses. Nevertheless, we feel that an error 
of order 10\% is acceptable considering the advantage of working with
analytical expressions. 

\begin{table}[t!]
\begin{tabular}{|r|r|c|c|c|}\hline
  {\mbox centrality}& $N_{\mbox{\tiny{part}}}$ & $R$ (fm)
  \\ \hline
  $ 0-10\%$ & 325.2 & 8.0\\ \hline
  $10-20\%$ & 234.6 & 7.3\\ \hline
  $20-30\%$ & 166.6 & 6.6\\ \hline
  $30-40\%$ & 114.2 & 5.9\\ \hline
  $40-50\%$ & 74.4  & 5.3\\ \hline
  $50-60\%$ & 45.5  & 4.6\\ \hline
  $60-70\%$ & 25.7  & 4.0\\ \hline
  $70-80\%$ & 13.4  & 3.4\\ \hline
  $80-92\%$ & 6.3   & 2.9\\ \hline
\end{tabular}
\caption{Parameters $N_{\mbox{\tiny{part}}}$ and $R$ for the description
  of the pion and proton spectra corresponding to different 
  centralities. The radii of the equivalent spherical region has been
  calculated according to the relation
  $R=R_0+C(N_{\mbox{\tiny{part}}}/2)^{1/3}$,  
  with $R_0=1$ fm and $C=1.28$.}  
\label{tab1}
\end{table}

We also notice that the description of the proton data for $p_t> 3$ GeV is
not as good. This can be understood recalling that for large $p_t$ the leading
particle production mechanism is the fragmentation of fast moving partons,
some of which fragment outside the fireball region and thus are not influenced
by the confining boundary that the rest of the particles experience within the
fireball and thus, that our description is not valid for these large $p_t$
particles.

Figure~\ref{fig2} shows the distributions for pions and protons for different
collision centralities. For the description of these data, we have fixed
the values of $T_{\mbox{\small{$p,\pi$}}}$ and
$\beta_{\mbox{\small{$p,\pi$}}}$ to the ones obtained from the most central
collisions analysis leaving the normalizations to be determined by the
minimization procedure. The size of the overlap region for peripheral
collisions has been determined from the number of participants
$N_{\mbox{\tiny{part}}}$ in the reaction~\cite{PHENIXBM} by a simple scaling
law for the size of the equivalent spherical region according to the relation
$R=R_0+C(N_{\mbox{\tiny{part}}}/2)^{1/3}$, with $R_0=1$ fm and
$C=1.28$ that gives $R=8$ fm for the most central region $(0-10\%)$
data. This relation is motivated by the similar one that gives the
radius of a nucleus in terms of the mass number. The value of $R_0$
tries to account for the finite size of the 
interaction region as the number of participants takes its
smallest value for the most peripheral collision, namely,
$N_{\mbox{\tiny{part}}}=2$.

The values for $R$ and 
$N_{\mbox{\tiny{part}}}$ are listed on Table~\ref{tab1}. We notice that the
pion data are well described for all centralities except at the lower end of
the spectra where a thermal calculation is expected to fail due to resonance
contamination. The proton data is well
described up to $p_t\sim 3$ GeV only up to centralities of order $40-50 \%$
from where the quality of the description decreases as the centrality of the
collisions decreases. We interpret the poor description of the proton data for
$p_t>3$ GeV for all centralities as an indication that the leading particle
production is not a kind of thermal parton recombination but instead the
fragmentation of fast moving partons. On the other hand, the failure to
describe proton data for centralities smaller that $40-50 \%$ could be
attributed to a different scaling of the effective size of the interaction
region with $N_{\mbox{\tiny{part}}}$ as compared to the one obeyed by the
pions or to the fact that for large impact parameters, the proton size becomes
comparable to the size of the interaction region. An analysis to explore these 
possibilities will be presented elsewhere.

\section{Particle Correlations}\label{IV}

In order to explore the space-time dimensions of the system created in
high-energy heavy-ion collisions, one typically looks at two-particle
correlation functions. In the present case, it is thus instructive to
look at this function to see whether the size scale that can be
extracted from a two-particle correlation is comparable to the
intrinsic scale dimension used in the model formulation. We carry out
the analysis for the two-pion correlation function. For the purposes
of this section, we closely follow Ref.~\cite{Zhang} to where we refer
the reader for details. 

Let $\psi_{nlm'} ({\mathbf p})$ represent the Fourier transform of
the wave function for the state with quantum numbers $n,l.m'$, namely
\be
   \psi_{nlm'} ({\mathbf p})=\int \frac{d^3r}{(2\pi)^3}e^{-i{\mathbf p 
   \cdot \mathbf r}}
   \psi_{nlm'} ({\mathbf r})\, .
   \label{FT}
\ee
With the normalization adopted in Eq.~(\ref{eq:solnew}), the one-pion
momentum distribution can be represented as  
\be
   P_1({\mathbf p})&\equiv&\frac{d^3N}{d^3p}\nonumber\\
   &=&\sum_{n,l,m'}
   2E_{nl}e^{-\bar{\gamma}E_{nl}/T}\psi_{nlm'}^*
   ({\mathbf p})\psi_{nlm'}({\mathbf p})\, .
   \label{distP1}
\ee 
Similarly, and under the assumption of a complete factorization of
the two-particle density matrix, the two pion momentum distribution can
be written as
\be
   P_2({\mathbf p_1},{\mathbf p_2})&\equiv&\frac{d^6N}{d^3p_1d^3p_2}
   \nonumber\\
   &=&P_1({\mathbf p_1})P_1({\mathbf p_2}) \nonumber\\
   &+&\left|\sum_{nlm'}
   2E_{nl}e^{-\bar{\gamma}E_{nl}/T}\psi_{nlm'}^*
   ({\mathbf p_1})\psi_{nlm'}({\mathbf p_2})\right|^2,\nonumber\\
   \label{distP2}
\ee
from where the two-pion correlation function $C_2$ can be written, 
in terms of $P_1$ and $P_2$, as  
\be
   C_2({\mathbf p_1},{\mathbf p_2})\!&=&\!
   \frac{P_2({\mathbf p_1},{\mathbf p_2})}
   {P_1({\mathbf p_1})P_1({\mathbf p_2})}.
   \label{correl}
\ee
No\-tice that as a con\-se\-quence of the fac\-tor\-iza\-tion 
assumption, the correlation function is such that 
$C_2({\mathbf p},{\mathbf p})=2$. 

For the spherically symmetric problem here described, the correlation
function depends on the magnitude, as well as on the angle between
the momenta of the two particles ${\bf p}_1$ and ${\bf p}_2$. We make
the change of variables to relative ${\bf q}={\bf p}_1-{\bf p}_2$ and
average ${\bf K}=({\bf p}_1+{\bf p}_2)/2$ momenta and also to the
angle between these last two vectors, $\theta$. The correlation
function thus becomes a function of $K=|{\bf K}|$, $q=|{\bf q}|$ and
$\theta$. In order to consider the contribution from pions with
different angles between their momenta, we 
average over $\theta$. Figure~3 shows $C_2(q)$ averaged over $\theta$
and for a fixed value $K=260$ MeV as a function of $q$ for $R=8$ fm,
$T=117$ MeV and $\beta=0.55$. The solid curve shows the corresponding
Gaussian fit.

\begin{figure}[t!] 
{\centering
\resizebox*{0.4\textwidth}
{0.34\textheight}{\includegraphics{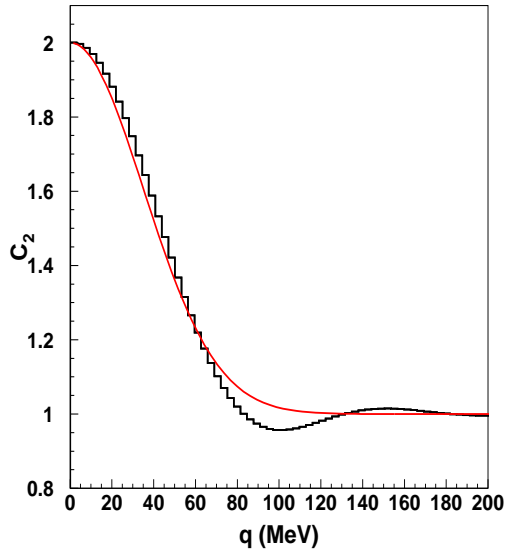}}
\par}
\caption{(Color online) $C_2(q)$ averaged over the angle between  ${\bf K}$ and
${\bf q}$ as a function of $q$ for $K=260$ MeV, $R=8$ fm, 
$T=117$ MeV and $\beta=0.55$. The histogram corresponds to the model
calculation whereas the solid curve shows the corresponding
Gaussian fit.}
\label{fig3}
\end{figure}

In order to extract the system's size $R_{\mbox{\tiny{eff}}}$ from
this function, we fit this curve to a Gaussian distribution of the
form
\be
   g(q)&=&1+\rho^2(q),\nonumber\\
   \rho(q)&=&\exp(-q^2R_{\mbox{\tiny{eff}}}^2/2).
   \label{gauss}
\ee
Figure~4 shows the behavior of $R_{\mbox{\tiny{eff}}}$ as a function
of $(N_{\mbox{\tiny{part}}})^{1/3}$ compared to measured values for
$R_{\mbox{\tiny{side}}}$~\cite{PHENIXcorr} from Au + Au collisions at
$\sqrt{s_{NN}}=200$ GeV. The lower solid curve corresponds to our model effective
radii whereas the upper solid curve is the model curve displaced by a
constant $\tilde{R}=0.8$ fm. The dashed curve corresponds to the best linear
fit to the data. We notice that the slope of our model curve is in
good agreement with the data. The fact that the intercept is different
from zero may indicate the existence of correlation scales in the data
that are not considered in our simple approach. 

Finally, recall that 
\be
   \rho(r)=\exp(-r^2/2R_{\mbox{\tiny{eff}}}^2)
   \label{spacedist}
\ee
is the spherically symmetric three-dimensional distribution in space
that gives rise to $\rho (q)$ upon Fourier transformation and that the
$R_{r.m.s.}$ radius from $\rho (r)$ is given by 
\be
   R_{r.m.s.}=\sqrt{3}R_{\mbox{\tiny{eff}}}.
   \label{rmsgauss}
\ee
On the other hand, for a rigid sphere, such as the distribution giving
rise to our model distribution, the $R_{r.m.s.}$ radius is given by
\be
   R_{r.m.s.}=\sqrt{3/5}R.
   \label{rmsours}
\ee
By equating these two r.m.s. radii, we see that in order to compare the
effective radius with the model one, the relation between them is given by
\be
   R=\sqrt{5}R_{\mbox{\tiny{eff}}}.
   \label{RandReff}
\ee

\begin{figure}[t!] 
{\centering
\resizebox*{0.4\textwidth}
{0.34\textheight}{\includegraphics{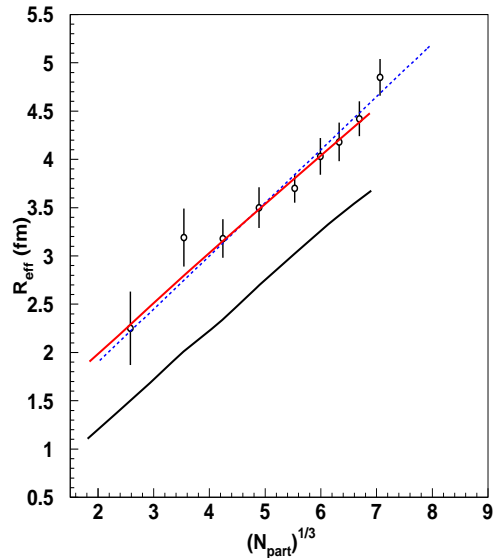}}
\par}
\caption{(Color online) $R_{\mbox{\tiny{eff}}}$ as a function
of $(N_{\mbox{\tiny{part}}})^{1/3}$ for $K=260$ MeV compared to measured values for
$R_{\mbox{\tiny{side}}}$ from Au + Au collisions at
$\sqrt{s_{NN}}=200$ GeV. The lower solid curve corresponds to the model effective
radii whereas the upper solid curve is the model curve displaced by a
constant $R_0=0.8$ fm. The dashed curve is the best linear
fit to the data}
\label{fig4}
\end{figure}

\section{Conclusions}\label{concl}

In conclusion, we have shown that by considering the finite size of the
interaction region and a simple scaling law of this with the number of
participants in the collision, it is possible to achieve a good description of
mid-rapidity pion and proton data in Au + Au collisions at RHIC with common
values of temperature and transverse expansion velocity. For central
collisions, the proton to pion ratio is also well described and its behavior
can be attributed to the strength of the radial flow achieved in
RHIC. By performing a two-particle correlation analysis and comparing
to data for $R_{\mbox{\tiny{side}}}$ as a function of
$(N_{\mbox{\tiny{part}}})^{1/3}$ from Au + Au collisions at
$\sqrt{s_{NN}}=200$ GeV, we see that the scaling law found from the single
particle spectra analysis is in good agreement provided that we
displace our model curve by a constant $\tilde{R}=0.8$ fm and we
speculate that this signals the existence of an extra correlation
length in data that is not accounted for in our simple model.

We should stress that the spherical symmetry assumed throughout
can be thought of as a theoretical tool rather than as a
realistic approximation to the actual collision geometry for the
highest RHIC energies. Our intention is to provide a working model
with a high degree of symmetry that can be better controlled in an
actual calculation. The same is true for the treatment of the shape of
the region for non-central collisions which 
one knows, from elliptic flow analyses, that partially retains the
original almond shape of the overlap region in the collision. The
sizes referred to in this way reflect characteristic sizes rather
than actual spherical radii. Although modifying the geometry used for
the calculation will certainly give rise to a different set of quantum
states, the bulk of the effect will remain since the physics that it
captures is the Heisenberg uncertainty principle whereby restricting
the size of the region to become finite, the momentum states become
broader. As for the use of stationary states, we point out that
although the freeze-out volume is 
reached with a large expansion velocity, the transition from a
strongly interacting system to a free gas is rapid and what matters is
the distribution right before this transition and therefore the length
scale associated to it.

It is important to emphasize that a description of the transverse
distributions for different impact parameters can be done by
considering a varying freeze-out temperature and radial velocity but
the lesson to be learned from the present analysis is that this
variation can be tempered and/or even avoided by considering the
finite size of the interaction region. 

We point out that some hydrodynamical models without finite size effects have
been able to give similarly good descriptions of data up to $p_t$ of order
2-2.5 GeV~\cite{Schnedermann} at the expense of introducing a large amount of
parameters. What we have shown here is that it is also
possible to achieve the same quality of description including a basic property
of quantum systems often neglected, that is the fact that in high energy
reactions, particles are produced in small space-time regions. While doing
this and in this first step approach, we made use of approximations to
render the calculations tractable, one of such is the treatment of the gamma
Lorentz factor in terms of an average one. The relaxation of these
approximations is a natural step forward and we will report on the progress of
this work elsewhere.

The authors thank G. Paic for his valuable comments and
suggestions. Support for this work has been received by PAPIIT-UNAM
under grant number IN107105  and CONACyT under grant numbers 40025-F
and bilateral agreement CONACyT-CNPq J200.556/2004 and 491227/2004-3.

\end{document}